\documentclass[twocolumn,amsmath,amssymb,10pt,superscriptaddress,a4paper,letterpaper,fleqn]{revtex4-1}
\usepackage{amssymb}
\usepackage{epsfig}
\usepackage{graphicx}
\usepackage{dcolumn}
\usepackage{array}
\usepackage{bm}
\usepackage{fancyheadings}
\usepackage{longtable}
\usepackage{multirow}
\usepackage{float}
\usepackage{afterpage}  
\usepackage{color}

\setlongtables
\usepackage[breaklinks=true,linkbordercolor={1 1 1}]{hyperref}

\begin{document}
\setcounter{page}{1}

\title{
\qquad \\ \qquad \\ \qquad \\  \qquad \\  \qquad \\ \qquad \\ 
Status report on the analysis of inelastic neutron scattering from carbon, iron, yttrium and lead at 96 MeV}

\author{C. Gustavsson}
\email[Corresponding author, electronic address:\\ ]{cecilia.gustavsson@physics.uu.se}
\affiliation{Dept. of Physics and Astronomy, Box 516, SE-751 20 Uppsala, Sweden}

\author{C. Hellesen}
\affiliation{Dept. of Physics and Astronomy, Box 516, SE-751 20 Uppsala, Sweden}

\author{S. Pomp} 
\affiliation{Dept. of Physics and Astronomy, Box 516, SE-751 20 Uppsala, Sweden} 

\author{A. \"{O}hrn} 
\affiliation{Westinghouse AB, V{\"a}ster{\aa}s, Sweden}

\author{J. Blomgren} 
\affiliation{Vattenfall AB, Stockholm, Sweden}

\author{U. Tippawan} 
\affiliation{Fast Neutron Research Facility, Chiang Mai University, Thailand}

\date{\today} 

\begin{abstract}

{This work is part of an effort to provide more experimental data for the (n,n'x) reaction. 
The experiments were carried out at The Svedberg Laboratory 
in Uppsala, Sweden, at the quasi-mono-energetic neutron beam of 96 MeV, before the facility was upgraded in 2004. 
Using an extended data analysis of data primarily intended for 
measuring elastic neutron scattering only, it was found to be possible to extract information on the inelastic scattering 
from several nuclei. 
In the preliminary data analysis, an iterative forward-folding technique was applied, in which a physically reasonable trial 
spectrum was folded with the response 
function of the detector system and the output was compared to the experimental data. As a result, double-differential cross 
sections and angular distributions of inelastic neutron scattering from $^{12}$C, $^{56}$Fe, $^{89}$Y and $^{208}$Pb could be obtained. 
In this paper, a status update on the efforts to improve the description of the detector response function is given.
}
\end{abstract}
\maketitle


\rhead{C.Gustavsson  \textit{et al.}}
\lfoot{}
\rfoot{}
\renewcommand{\footrulewidth}{0.4pt}

\section{INTRODUCTION}

Inelastic neutron scattering (n,n'x) is a pre-equilibrium reaction, i.e. an intermediate process between direct and compound nuclear reactions. 
The (n,n'x) reaction channel is much weaker than the elastic scattering (n,n) channel and there are very few data sets reported for 
inelastic neutron 
scattering at intermediate neutron energy (20-200 MeV).

In several applications involving neutrons, i.e. various nuclear power systems including accelerator-driven systems, the (n,n'x) reaction 
plays an important role. Firstly, it increases the neutron flux because after inelastic neutron emission, the 
residual nucleus in many cases emits a neutron after statistical decay (compound emission).
Moreover, in these intermediate-energy (n,n'x) experiments, only one neutron is detected, and in many instances the 
undetected particle (x) is a second neutron. Secondly, the (n,n'x) reaction lowers the neutron energies, which is of large importance 
for neutronics, shielding and radiation damage. At present, the (n,n'x) reaction is among the least studied 
of the intermediate-energy neutron reactions taking place in today's and future nuclear power reactors.

\section{EXPERIMENT}

The original experiments analysed in the work presented here, were performed at the neutron beam facility at The Svedberg
Laboratory in Uppsala, Sweden, during the years 2000-2004. A detailed description of the 
neutron beam facility can be found in in Ref.~\cite{Klug02}. 

\begin{figure}[!htb]
\includegraphics[width=1.0\columnwidth]{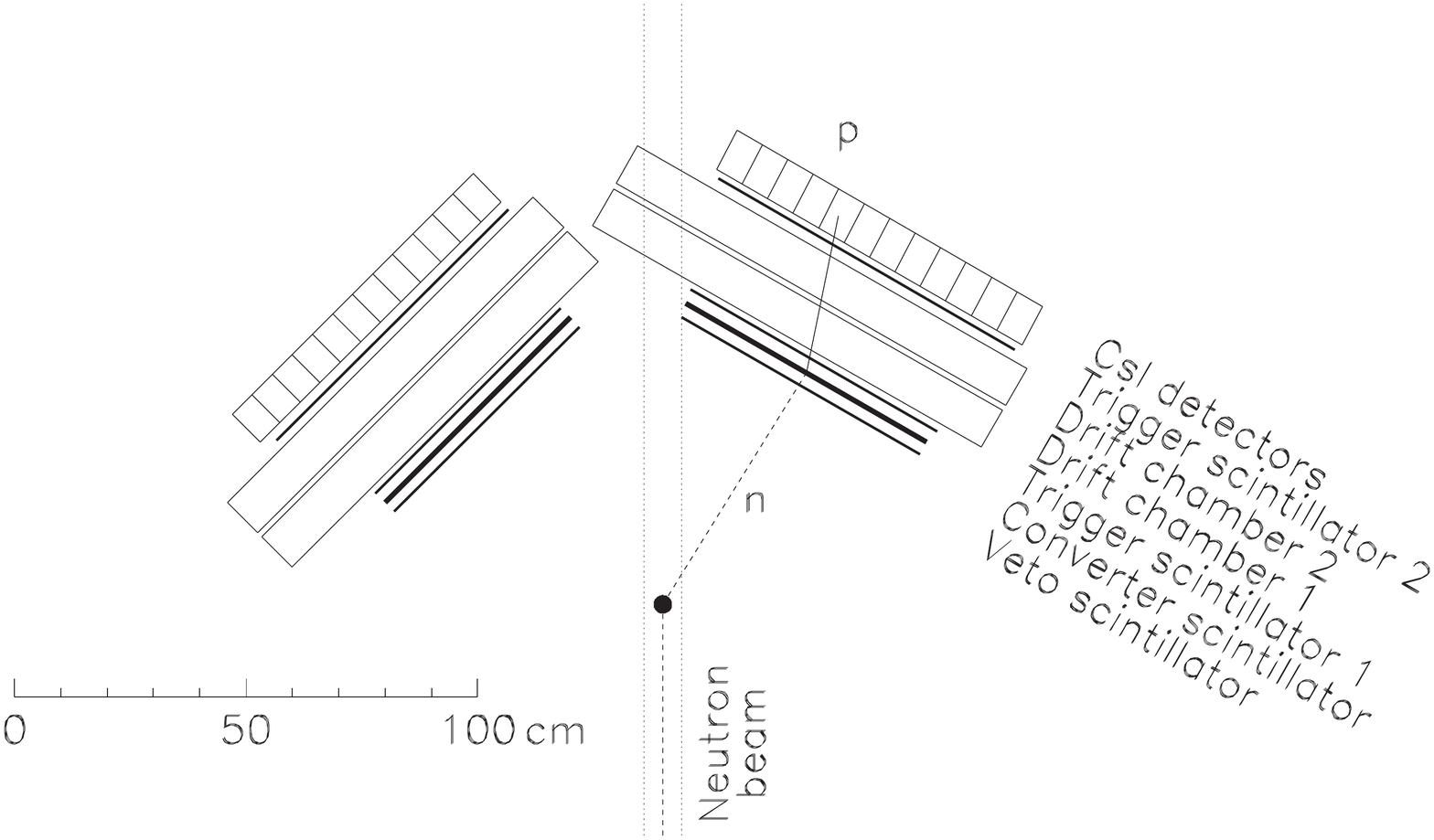}
\caption{A schematic view of the SCANDAL, SCAttered Nucleon Detection AssembLy, setup. 
Neutrons are entering from the bottom of the picture and 
scattered in the target before impinging on the setup.}
\label{scandal}
\end{figure}

The experimental setup SCANDAL (SCAttered Nucleon Detection AssembLy) was used to detect 
the scattered neutrons (see Fig.~\ref{scandal}). The detection of neutrons is based 
on conversion to protons in an active plastic scintillator and detection of the recoil protons in CsI crystals.
The setup consists of two identical arms placed on each side of the beam, 
covering the angular ranges 10-50 degrees and 30-70 degrees. Each arm has a 2 mm 
thick veto scintillator for fast rejection of charged particles, a 10 mm thick neutron-proton 
converter scintillator, a 2 mm thick $\Delta E$ plastic scintillator for triggering
and particle identification, 
two drift chambers for proton tracking, a second 2 mm thick $\Delta E$ plastic scintillator 
which is also part of the trigger, and an array of CsI detectors (12 crystals on each arm) for 
energy determination of the recoil protons produced in the converter.
The trigger is defined by a coincidence between the two trigger 
scintillators, with the front detector acting as a veto. It is also possible to run SCANDAL 
in proton mode, by including the veto detector in the trigger condition and thereby 
accepting charged particles. The total energy resolution of the individual CsI crystals 
is slightly different, and on average 3.7 MeV (FWHM) at 96 MeV, see Ref.~\cite{Klug02}. 

Details for the experiments discussed in this paper, can be found in Refs.~\cite{Mer06,Ohrn08,Klug03} 
where data on elastic neutron scattering from carbon (see Ref.~\cite{Mer06}), 
iron and yttrium (see Ref.~\cite{Ohrn08}) and lead (see Ref.~\cite{Klug03}) have been published.

\subsection{Detector response function}
\label{response}

The SCANDAL setup has a non-trivial response to incoming neutrons. For a given incident 
neutron energy impinging on the setup, a proton recoil energy spectrum is produced.
Neutrons converted to protons via $np$ scattering, i.e., H(n,p) reactions, 
in the converter scintillator produce a single peak in the response spectrum, 
located at the highest possible energy, i.e.,
close to the energy of the incident neutron. In addition, a distribution of protons at 
lower energies is produced by $^{12}$C(n,px) reactions due to the carbon content in the 
converter. This reaction has been studied experimentally at the same 
energy~\cite{Ols93}. These $^{12}$C(n,px) data cover, 
however, only excitation energies up to 40 MeV. 
Given that the $^{12}$C(n,p)$^{12}$N reaction has a Q-value of -12.6 MeV, 
this means that data for the carbon conversion contribution to the SCANDAL response 
are available up to slightly above 50 MeV excitation energy.

Another contribution to the response comes from the low-energy tail in the incident 
neutron beam. The $^{7}$Li(p,n) reaction has been measured at nearby 
energies, see Refs.~\cite{Byr89,Con90}, allowing its contribution to the response to be estimated 
by parameterizations, see Ref.~\cite{Pro02}. Finally, 
the detected spectrum is slightly distorted by the response to protons in the individual CsI crystals. 

With knowledge of all these contributions, the full response of the SCANDAL setup could 
be obtained. Such a response function has been studied experimentally in Ref.~\cite{Klug02}, 
and it was found that it describes data successfully. A pre-requisite is, however, 
that no time-of-flight cut is applied on the incident neutrons, because the exact 
effect of such a cut cannot be well described analytically. To circumvent this problem, a choice was made to determine the 
response function by measuring it at small neutron scattering angles.

In the previous works where preliminary results for the (n,n'x) reaction have been presented by our group, see Refs.
~\cite{Ohrn11,Gust12}, an experimentally measured response function has been used. It was obtained by measuring the (n,n) 
spectrum at small neutron scattering angles (9 degrees, which is the smallest accessible angle in SCANDAL) 
and that measurement was taken to represent the detector response function. The reason why 
this is possible is that because at small angles, elastic scattering dominates heavily meaning that the 
neutron spectrum hitting the SCANDAL setup is very similar to the incident neutron spectrum impinging onto the scattering target. For 
this approximation to be useful at all neutron energies, a method was adopted in which the response at lower energies 
was considered to have the same shape as the one at full energy. Therefore, the spectrum obtained at 
96 MeV incident energy was shifted in steps of 1 MeV in the subsequent analysis. 
This strategy was based on studies of the energy dependence of $np$ scattering and the $^{12}$C(n,p) reaction. 
It was found that the ratio of these two cross sections stays fairly constant over the 
relevant energy interval.

This solution to the problem of determining the detector response function is however an estimate and in order to 
improve the data analysis we have decided to perform a \textsc{GEANT4} (see Ref.~\cite{geant}) 
simulation of the response function and a subsequent re-analysis of data. In this way we hope to be 
able to present final data on the (n,n'x) reaction for carbon, iron, yttrium and lead.

\section{ANALYSIS PROCEDURE}

\begin{figure}[!htb]
\includegraphics[width=1.0\columnwidth]{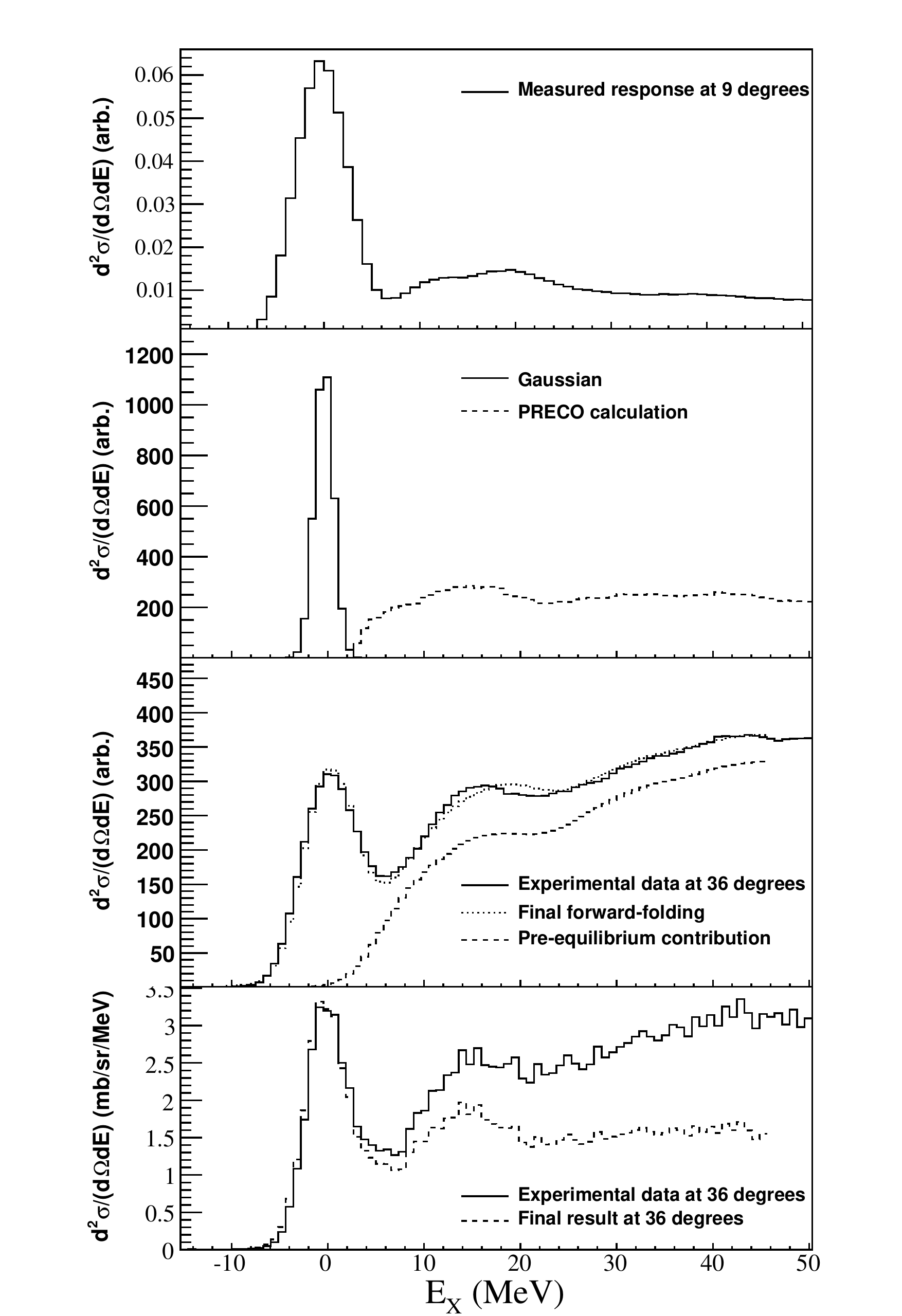}
\caption{Steps in the analysis procedure. See the text for details.}
\label{analysis}
\end{figure}

Once the detector response function is determined, the analysis of the data can be carried out using a forward-folding technique.
This means that a trial spectrum is folded with the experimental response. In panel 1 of Fig.~\ref{analysis}, representing the analysis of 
$^{56}$Fe at 36 degrees, a preliminary detector 
response function is shown, based on measurement of the (n,n) spectrum at small angles as described in Sect.~\ref{response}. Panel 2 in the same 
figure shows a trial spectrum consisting of a Gaussian representing the ground state (elastic scattering) and a continuum 
predicting the inelastic scattering. The continuum was modelled using the \textsc{preco} code see Refs.~\cite{Kal95,Kal04,Kal05,Kal06}. 

In the first step of the analysis, the trial spectrum is folded with the detector response function and compared to the experimental data. 
The difference between the output of the forward-folded trial spectrum and the experimental data is then used to modify
the input to the next forward-folding. This is an iterative method and it has been found that typically three iterations are required
to obtain a stable output. An example is shown in panel 3 of Fig.~\ref{analysis}, where the solid histogram represents the experimental data, the 
dotted histogram is the output from the forward-folding and the dashed histogram is the contribution from the continuum part of the trial spectrum.

When the final forward-folded spectrum has been obtained, a ratio between the input and the output of the 
total forward-folding procedure is used to establish a correction with which the measured experimental spectrum can be multiplied 
bin-by-bin. The result of this correction is shown in panel 4 of Fig.~\ref{analysis}. 
The upper histogram (solid) shows the experimental data before corrections and the lower histogram (dashed) after 
corrections. 

For the final data, we also apply corrections for the energy dependence of the $np$ cross section and for 
variations in proton detection efficiencies
in different CsI crystals. Corrections for attenuation and multiple neutron scattering in the scattering targets are also performed, 
employing two different techniques, as described in Ref.~\cite{Ohrn08}. Both corrections are based on Monte Carlo codes. 
\vspace{4 mm}

\section{CONCLUSIONS AND OUTLOOK}

To finalize this work, an ultimate detector response function must be determined. With the aid of a \textsc{GEANT4} simulation taking into account 
the geometry and constitution of the SCANDAL setup, as well as the spectrum of the incoming neutrons, the response to all neutron energies can be 
obtained. 
Once this work has either confirmed or altered our preliminary analysis we can publish our final results for neutron inelastic 
scattering from carbon, 
iron, yttrium and lead and compare them to relevant models and available data.

\begin{acknowledgments}

This work was supported by 
the Swedish Nuclear Fuel and Waste Management Company, the Swedish 
Nuclear Power Inspectorate, Ringhals AB, Forsmark AB, the Swedish Defence Research 
Agency, the Swedish Research Council and the European Council.

\end{acknowledgments}

\end{document}